\def\beq{ \begin{equation}}
\def\eeq{ \end{equation}}
\def\ba{ \begin{eqnarray}}
\def\ea{ \end{eqnarray}}
\def\pound {{\cal L}}
\def\half {{1\over 2}}
\documentstyle[aps,preprint,12pt]{revtex}
\tightenlines
\title{ Cosmological long wavelength perturbations}

\author{ W. Unruh}
\address{ Program in Cosmology and Gravity of CIAR\\Dept 
Physics and Astronomy\\University 
of B.C.\\ Vancouver, Canada V6T 1Z1}
\begin{document}
\maketitle

\begin{abstract}
This paper presents an exact solution to the long wavelength perturbations
 for the scalar modes and for a scalar field theory with arbitrary potential. 
Locally these modes are coordinate transformations of the homogeneous background
 solutions (although non-locally they are not). These solutions are then used
 to discuss a couple of recent papers in which such perturbations play 
a role.
 Abramo, Brandenberger, and Mukhanov have recently argued that long wavelength 
perturbations have the effect of driving the cosmological constant to zero 
if the higher order perturbation equation are examined. I argue that this 
effect is invisible to any local observer, and thus does not constitute 
a relaxation of the cosmological constant in the normal sense of the term.

 Grishchuk has argued that the standard lore on the strength of the perturbations 
at the end of inflation is wrong.  I discuss the disagreement in light of the
 exact long wavelength solutions, and emphasize the importance of the initial
 conditions in resolving the disagreement.
\end{abstract}

In the inflationary models for the growth of the universe, the physics 
of long 
wavelength perturbations (wavelengths which are much longer the the Hubble
 radius) play an important role. It is the behaviour of the perturbations 
during
 this time period which determine the effect of the perturbations on the 
present
 structure of the universe.  Recently, a couple of papers have discussed 
these
 long wavelength perturbations in different contexts. 
Abramo, Brandenberger, and Mukhanov \cite{ABM} have argued that higher
 order corrections 
to Einstein's equations for long wavelength perturbations have the effect 
of creating a negative cosmological 
constant which reduces the effective cosmological constant for the space-time. 
They also speculate that this   could be a mechanism to drive the actual 
cosmological 
constant to near zero (or in particular to near the current mass density 
of the universe). On the other hand, Grishchuk\cite{G,Greply} has  claimed 
that the standard calculations for the size of the density perturbations 
at the
 end of inflation are wrong. Various authors \cite{MD,MS} have claimed 
that his 
treatment of the behaviour of the long wavelength perturbations is wrong.

In this paper, I derive an exact expression for these long wavelength perturbations
 in a system in which the dominant gravitating matter is a scalar field 
$\Psi$ with 
an arbitrary potential $V(\Psi)$. I note in passing that these solutions 
therefor 
apply without much change to the evolution with an arbitrary perfect isotropic 
fluid, under the usual identification that 
\ba
&&\rho+p={1\over 2} g^{\mu\nu}\Psi_{,\mu}\Psi_{\nu} \\
&&u_\mu = {\Psi_\mu  \over \sqrt{\rho+p}} \\
&&\rho-p =V
\ea

\section{ Gauge Invariance vs Gauge Fixing}

Let me first discuss the ABM paper.
I was confused by their result that the long wavelengths could renormalise 
the cosmological constant since the effect is caused by the averaged 
 energy momentum tensor of modes whose 
wavelength is much larger than the Hubble's radius at any time of interest. 
``How could such long wavelength modes   affect the cosmological constant 
as 
seen by an observer who can measure the cosmological constant only over 
a region smaller than his own Hubble radius?" I would have expected  
modes, whose wavelength is much much larger than the Hubble radius, to 
look like
a   homogeneous universe to a local observer.  
  
In addition, as they point out, the   effective stress energy tensor of 
lower order
 perturbations
can be   coordinate dependent ( called gauge dependent if one 
is looking at small 
coordinate transformations). Their approach to this problem was to write 
the equations in terms of "gauge independent" variables. However, as I will argue in the 
following,   there is no difference between such a "gauge invariant" approach, 
and an approach which fixes the gauge in some way. Any variable in a gauge 
fixed formalism is completely equivalent to some  gauge independent variable, 
and vice-versa. 
In particular, this means that, even though the effective stress energy 
tensor is not independent
on gauge transformations in the gauge invariant approach, it is dependent 
on which set of gauge 
invariant variables one chooses. In the gauge fixing approach, the same 
problem arises in that the 
effective stress energy tensor is dependent on which gauge fixing choice 
one makes.
 
Let me review their approach.  Instead of writing  Einstein's equations
in terms of the free metric and matter perturbations, $\delta g_{\mu\nu},
~\delta\Psi$,
 they choose
a set of vector fields $X^\mu$ created out of the metric and field variables, 
(i.e., $X^\mu=X^\mu(\delta g_{\lambda\sigma},\delta\Psi)$ ), 
and defined new 
gauge invariant perturbations 
\beq
\delta\bar g_{\mu\nu}= e^{\pound_X}(g_{(0)\mu\nu}+\delta g_{\mu\nu}) -g_{(0)\mu\nu}
\eeq
where $\pound_X$ is the Lee derivative with respect to $X$. A similar definition 
applies to 
$\delta\bar\psi$. $X$ is chosen so that the barred quantities are invariant under a
 (first order) gauge transformation. 

Given the free first order  metric perturbations written as
\beq
\delta g_{\mu\nu}=\left( \begin{array}{lc}
                            \Phi +S^i_{;i} ~~~~~~~~~& S_i+B,i \\
                            S_i+B,i &   \phi g_{(0)ij} +Q_{|ij}+R_{i|j}+h_{ij}
                         \end{array}
\right)
\eeq
where $\phi, B, {\rm and }Q$ represent the scalar perturbations,  the
transverse vector fields,$S_i,~R_i$ represent the 
 vector perturbations, and the transverse-traceless tensor $h_{ij}$
 represents the gravitational wave perturbations.  $g_{(0)\mu\nu},~ \Psi_0 $ 
are
 assumed to be  a background flat-space homogeneous   solution. 
In the 
following, I will concentrate on the scalar perturbations alone ($S_i,R_i$, 
and $h_{ij}$
 will be neglected.

The vectors $X^\mu$ are now chosen to be the infinitesimal coordinate transformations
 required to reduce the above general metric to some fixed   form. For 
example, one 
choice (used by ABM) is
\beq
X_\mu= [ \dot  Q-2HQ -B, ~~-\partial_i Q  ]
\eeq
This is exactly the infinitesimal coordinate transformation required to 
make the $B$ and
 $Q$ terms in the perturbation metric equal to zero, leaving only the
  diagonal $\phi,~\Phi$ terms 
non-zero. But the equations of motion for these terms are also exactly the equations 
of motion for the gauge fixed metric, where the gauge fixed $B$ and $Q$ are set equal 
to zero.
Although ABM mention a few such possible choices for the vectors $X^{\mu}$, 
there
 are a large number which they do not mention because they 
restricted themselves to use only the local metric variables
 in defining $X^\mu$, while the use of the matter variables and of non-local metric 
variables would allow an expanded set of possibilities. In fact, for any 
gauge fixing,
 one can find a vector $X^{\mu}$ which would implement that gauge fixing. 
For example,  if we choose
\beq
X^{\mu}= \left[ -{\delta\Psi\over \dot \Psi_0}, 
-\partial_i \int(B-{\delta\psi\over\dot \Psi_0}){dt\over a^2} 
\right]
\eeq
 the gauge invariant scalar field 
perturbations,$\delta\bar\psi$, will be zero, as will the off diagonal temporal terms in 
the metric,$\bar B$. This choice of vector 
 corresponds to the gauge fixing for the first order perturbations of the  
scalar field and scalar metric of $\delta\psi=0$, and $\delta g_{0i}=0$.
The choice
\beq
X^{\mu}= \left[-\int\phi dt, 
-\partial_i(\int{1\over a^2}\left(\int \phi dt\right)  -B)dt\right]
\eeq
on the other hand puts the gauge invariant metric into the synchronous 
gauge. One problem with using such non-local gauge transformation is  that
 because of the integration ``constants" (actually functions 
of the spatial variables) in $X$, these choices still retain some residual 
gauge freedom, corresponding to those integration ``constants".  

The procedure which they follow is to transform their original general metric by
means of these gauge transformations into what they call a gauge-invariant
metric. However, this is simply the procedure of making an appropriate
coordinate (gauge) transformation so as to put the metric into the appropriate
gauge fixed form. They then calculate the second order fluctuations in this
gauge fixed form, and obtain a result which they then claim to gauge invariant.
It is, in the sense that the results have been calculated in a specific gauge,
and thus that specific gauge allows no further gauge transformations. (or to put
it in another language, the gauge transformation is undone by their reduction of
the metric).

The gauge invariant approach is thus identical to the gauge fixing approach
and any problems or advantages of one are also problems and advantages
of the other.

\section{Long Wavelength Solutions}
To further the discussion, let me now solve the perturbation equations in the 
long wavelength limit. I will, with 
ABM, choose the longitudinal gauge fixing, which gives the following  equations.
The linear metric in this gauge is given by
\beq
ds^2=(1+2\Phi) dt^2 - a^2(t) (1-2\phi) (dx^2+dy^2+dz^2)
\eeq
where $\phi(t,x,y,z), \Phi(t,x,y,z)$ are the first order perturbation of 
the metric, and 
$\psi$ is the first order perturbation of the scalar field $\Psi$ (called
$\delta Psi$ above).  The first
 order Einstein's equation $G_{ij}-T_{ij}=0$ for $i\neq 
j$ gives
\beq
\partial_i\partial_j(\phi-\Phi)=0,
\eeq
which by rotational invariance gives
\beq
(\partial_i\partial_j-{1\over 3}\nabla^2\delta_{ij})(\phi-\Phi)=0
\eeq
for all $i,~~j$
The solution is  
\beq
\Phi=\phi +\kappa_i(t)x^i+\sigma(t)(x^2+y^2+z^2)+\nu(t)
\eeq
 The term proportional to $\sigma$ corresponded to a perturbation toward one of the 
spherical or hyperbolic homogeneous space-times. 
The $\nu$ and $\kappa_i$ terms can  be removed by infinitesimal   coordinate transformations. I will 
ignore these possibilities here.  

Einstein's equations are
\beq
E_{\mu\nu}=G_{\mu\nu}-T_{\mu\nu}=0
\eeq
Selecting the two constraints,  we have
\beq
0=E^{(1)}_{tt}(\phi,\psi)=  -6H\dot \phi -V\phi-{1\over 2}V'\psi
 -\half\dot \Psi_0\dot \psi+{2\over 
a^2}\nabla^2\phi
\eeq
\beq
0=E^{1}_{ti}=-{1\over a} \partial_i\left(2\partial_t (a\phi)
 + \half a\partial_t(\Psi_0)\psi\right)
\eeq
The remaining (spatial diagonal) terms,  $E^{1}_{ii}$, are obtainable from these via  the 
conservation equations $E^{\mu\nu}~_{;\nu}=0$.

In addition the linearized equation for the scalar field is
\beq
P^1=\ddot\psi+3H\dot \psi -\nabla^2\psi + V''(\Psi_0)\psi 
-4\dot \Psi_0 \dot \phi-6H\dot \Psi_0\phi 
- 2 \ddot\Psi_0\phi=0
\eeq
In the above,  $\Psi_0(t)$ is the lowest order solution for the scalar field, $a(t)$ 
is the lowest order scale factor, $V(Psi)$ is the potential for the scalar 
field, and $H=\dot  a/a$ is the background Hubble  constant. These
variables obey
\beq
6H^2= {\half}\dot \Psi_0^2+V
\eeq
\beq
{\half}\dot \Psi_0^2= -2\dot  H
\eeq

In the long wavelength limit, where we neglect the term $\nabla^2\phi$, 
the equation $E^{1}_{tt}=0$ and the linearized equation  for the scalar 
field $P^{1}=0$
 are also  linearized equations for   homogeneous perturbations of the universe. 
  But at least two of those homogeneous solutions 
are 
simply coordinate transformations. These two correspond to  multiplying the spatial variables
 by a constant $\beta$ and the translating the time by an amount $\lambda(t)$. Ie, 
the 
new coordinates are 
\ba
\bar t=t-\lambda(t)\\
\bar x^i= (1+\beta)x^i
\ea
The metric now becomes
\beq
ds^2= (1+2\dot \lambda)d\bar t^2 
- a(\bar t )^2(1-2\beta+2H\lambda)\delta_{ij}d\bar x^i d\bar x^j
\eeq
Demanding that these transformations leave the metric  in the longitudinal gauge  
with $\phi=\Phi$ we get
\beq
\dot \lambda=\beta- H\lambda
\eeq
which we can easily solve to give
\beq
\lambda(t)= {\beta \int a dt +\epsilon  \over a}
\eeq

Thus,   
\ba
&&\phi= \partial_t \left( {\beta \int a dt +\epsilon  \over a}\right)f(x^i)=
(-H {\beta \int a dt +\epsilon  \over a} +\beta)f(x^i) \\
&&\psi= \dot \Psi_0 {\beta \int a dt +\epsilon  \over a}f(x^i)  
\ea
must be solutions to $E^1_{tt}=0$ and $ P^1=0 $ if we neglect the 
terms
 proportional to $\nabla^2$ (ie in the long wavelength limit).
The only remaining equation is the $E^1_{t i}=0$ equation. But again, 
substituting our trial  solution into that equation clearly satisfies it.
 
Since the equation  $P^1=0$ is a second order temporal equation in $\psi$, 
and the $E_{tt}=0$ 
is a first temporal order equation in $\phi$, we expect three linearly 
independent solutions 
to these two coupled linear equations. The above are two of them. However,  
 solving $E_{ti}=0$  for $\psi$ and substituting into $E^1_{tt}=0$, we obtain
 a single second order equation in $\phi$. Since the $E^1_{tt}=0$ equation
 uniquely determines $\psi$ if $\phi$ is known, the linearized long wavelength
 equations have only two solutions, and we already know two of them.
 The above two solutions are the only solutions to the long wavelength
 equations. 

Note that without the $E^1_{ti}=0$ equation, there would have been 
three
 solutions. In the infinite wavelength limit, the $E^1_{ti}=0$ equation is 
satisfied identically (because of the spatial derivative in that
equation), and thus there exist three linearly independent solutions to the linearized equations. 
As I demonstrated above,  two of them correspond  
to coordinate transformations of the 
original background solution, while the third would correspond to true 
a physical
 mode. It is thus 
interesting that both physical long wavelength solutions go to coordinate 
transformations in the infinite wavelength limit, while the actual physical
 infinite wavelength solution has no correspondence for large but finite 
wavelengths.
 
Of course, having solved the problem in one gauge, one can determine the 
solution
 in any gauge.
For interests sake, and because Grishchuk uses that gauge, let us look at 
the
 solutions in synchronous gauge. Since we have the solutions in longitudinal 
gauge, the problem in synchronous gauge can be obtained by simply making
 the appropriate gauge transformation. The gauge transformation vector to
 go from longitudinal to synchronous gauge is given by
\beq
V^\mu = \left[ -\int\phi(t,x,y,z)dt, ~~\partial_i \int{\int \phi dt \over 
a^2}dt \right]
\eeq

The transformed metric is
\ba
&&\delta g_{(1)tt}=\delta g_{(1)ti}=0 \\
&&\delta g_{(1)ij}= 2a^2\left(  (\phi +H \int \phi dt) \delta_{ij} 
+\partial_i\partial_j \int{\int \phi dt \over a^2}\right)
\ea
Inserting the  known solution for $\phi$, i.e.,
\beq
\phi=f(x,y,z) \partial_t\left({\epsilon+\beta\int a dt\over a}\right)
\eeq
we obtain
\ba
\delta g_{(1)ij}=&& 2 a^2\left( (\beta f(x,y,z) +\kappa(x,y,z) H )\delta_{ij}+
\partial_i\partial_j f(x,y,z) \int {\epsilon +\beta\int a ~dt \over a^3} 
dt \right. \\
 &&\left. + \partial_i\partial_j (\kappa(x,y,z) \int a^{-2} dt + \gamma(x,y,z))\right)
\ea
$\kappa$ and $\gamma$ are temporal integration ``constant" spatial
 functions. Both correspond to 
gauge transformations, illustrating the well known feature of the synchronous
 gauge that it does not completely specify the gauge. These terms could 
be
 removed by a gauge transformation which leaves the system in synchronous
 gauge. Note that the part of the metric which is the $\bar \phi$ term 
 (whose contribution to the metric in synchronous gauge is proportional to $\delta_{ij}$) 
 contains only the $\beta$ parameter of the two parameter family of physical
 solutions.  The other physical solution, given by the terms which depend 
on 
$\epsilon$, occurs only in the spatial derivative $\bar Q$ terms. Any second
 order equation for $\bar \phi$ would therefor pick up only one of the 
physical modes.   Fortunately the solution which depends on $\epsilon$ 
is one which dies out
 at long times in an expanding universe, and as a result their neglect 
would not
 be of importance in the late stages. Note that $\bar\phi$ is  physically 
constant (up to a gauge transformation) for all times, agreeing with the contention
 of Grishchuk (where this   term is given the name of $h$).

Examining  the long-wavelength perturbation in the scalar field, we find
in synchronous gauge that 
\beq
\psi= \dot \Psi_0(t) \kappa(x,y,z)
\eeq
The only term remaining in the scalar field perturbation is a pure 
gauge term.
 Ie, synchronous gauge in the long wavelength limit  is also a gauge in 
which 
 the first order scalar field perturbations are zero, modulo a gauge transformation.

\subsection{Grishchuk's Criticism}
Let me now comment on the controversy between Grishchuk and
the rest of the community. As has been emphasized  by Grishchuk, who operates
 in synchronous gauge, the "growing mode" in synchronous gauge is constant 
for
 long wavelength modes. In the above, this is just the constant $\beta$. 
He then
 argues that this indicates that there is no amplification of the scalar 
modes in inflation while
 the modes are outside the Hubble radius (ie long wavelength). Since in 
this regime
 the scalar field (matter) perturbations are  zero,
 do not display any ``amplification" either. 
His
 conclusion is that this indicates that the scalar modes are of the same 
size at
 the end of inflation (and in fact when they reenter the Hubble radius after reheating) 
as they
 were at earlier times when those modes left the Hubble radius (when 
the physical wavelength
 first became larger than the Hubble radius). Since the gravity wave modes also
 do not grow during inflation, this suggests to Grishchuk that both scalar 
and gravity wave modes should have roughly the same size at the end of inflation. 
This is in direct contrast with the standard lore, that in general the 
gravity
 wave modes are much smaller than the scalar modes.

On the other hand, if one works in the longitudinal gauge, the growing mode of the gravitational
 metric parameter  $\phi$ goes as 
$\beta H \int a dt /a-\beta$, which is (modulo decaying modes)  identically
 zero  if the expansion is exactly exponential, and is $\beta/\mu$ if 
the
 expansion goes as a power law $a(t)\propto t^\mu$. Furthermore the
 scalar field modes go as $\dot \Phi (\beta (t-t_0)/(\mu+1))$ for a power
 law increase in $a$. Since both $\dot \Psi$ and  $t$
 increase as inflation continues, these scalar field modes grow from
 the start of inflation (where $\dot \Psi$ and $t-t_0$  are small) till the end (where
 $\dot \Psi$ is large.). 

But of course neither of these statements mean
 anything  in themselves. It is clear that statements about the growth
 of perturbations are highly coordinate dependent statements, and
 thus contain no physics per se. The key point is that arguing whether
 or not a perturbation grows or does not grow is irrelevant if no statement
 is also made as to how large the perturbations were at the beginning of
 inflation. The whole argument between Grishchuk and others  can be
reduced
 to the contention on the one hand that it is $\psi$ in the longitudinal
 gauge which has a roughly vacuum quantum amplitude $1/\sqrt{H}$ (in units 
where $\hbar=1$) at the time when the modes cross the Hubble radius,
 while the other contends that it is the $h$ mode, the diagonal term in
 the synchronous perturbation,
\beq
ds^2= dt^2- (a^2(1-2h)\delta_{ij}+Q_{,i,j})dx^idx^j
\eeq
 which has that amplitude. (This view of the debate is weakened by
noting that in \cite{G}, Grishchuk seems to adopt the same initial
condition conventions as others). In short, it is crucial to decide how
one will quantize the modes at the earliest times. By an appropriate
choice any final result can be obtained.  

 In also seems to an outsider that energy has been wasted
 about the details of the long wavelength calculations-- Grishchuk contending 
the favourite ``gauge independent variable" of the one group $\zeta 
= {\dot \phi/H+\phi\over \dot  H}+\phi$ is zero in the long wavelength limit (it is not--it is just $\beta$
 in the above notation--see also \cite{Goetz},\cite{Caldwell}), while the others contend that he has not done
 his matching properly  as the equation of state of the background
 changes (his $h$, as shown above can be chosen to be constant at
 all times, no matter what the potential $V$ in scalar field terms, or
 no matter what the equation of state in perfect fluid terms. It is however
not clear that he has always chosen this gauge.\cite{MD})

Trying to decide what the initial 
conditions are for the perturbations in the very early universe is a thorny issue. One
 tactic is to declare that these perturbations must be ``vacuum" perturbations.
 However the justification for this stance is somewhat weak, since by
 assumption (crucial for inflation to work at all)
 at least a part of the system is very far from its lowest energy state.
 The scalar field must have a very large value, and a large non-zero energy for inflation 
to proceed.
 However, to assume that it is far from equilibrium, while all other degrees of freedom 
at at their minimum energy is worrisome. But this is not the place to try to
 examine this issue in any detail. Thus I will use the assumption that the
fluctuations are in some sort of  minimum energy state.
Energy is however a coordinate dependent quantity, and, because of the large non-equilibrium
 background field, there is a tight coupling between the gravitational 
degrees of freedom and the matter. One must therefor adopt a formalism 
which takes this into account. Fortunately, because of the background field, and the special 
coordinate system chosen for that field, one can define an energy for the fluctuations to
lowest order in the those fluctuation.
 
\section{Reduced Hamiltonian Action}
Let me therefor derive the reduced Hamiltonian for
 the scalar field perturbations in a flat FRW space-time 
 (see also Mukhanov and Anderegg\cite{MA} and Garriga et. al \cite{Sasaki}).
The Hamiltonian  action for metric perturbations is
\beq
S= \int (\pi^{ab}\dot \gamma_{ab} +\pi_\Psi \dot \Psi- N H_0 -N_b H^b )d^xdt
\eeq
where
\ba
&&H_0={1\over \sqrt{\gamma}} \pi^{ab}\pi^{cd} ( \gamma_{ac}\gamma_{cd}
-1/2\gamma_{ab}\gamma_{cd}) - \sqrt{\gamma} R+ {1\over \sqrt{\gamma}}\pi_\Psi^2 
+\gamma^{ab}\Psi_{,a}\Psi_{,b} +V(\Psi)\\
&&H^a=\sqrt{\gamma} D_b { \pi^{ab}\over \sqrt{\gamma}} + \pi_\Psi \Psi_{,b}\gamma^{ab}
\ea
where $\gamma=det(\gamma_{ab})$ and $D_a$ is the covariant derivative with 
respect to the spatial metric $\gamma_{ab}$. I will now write this action
 in terms of the first order perturbations for the scalar modes. Writing 
\ba
&&\Psi=\Psi_0(t) +e\psi\\
&&\pi_\Psi=a^3\dot \Psi_0 +e \pi_\psi\\
&&\gamma_{ab}= a^2(1+2\phi)\delta_{ab} +Q_{,a,b}\\
&&\pi^{ab}=-2Ha\delta^{ab} +e \left( ({1\over 4a^2}P_\phi
 + \half  \nabla^{-2}P_Q)\delta^{ab} + {1\over a^2}(- 3\nabla^{-2} P_Q
-{1\over a^2}P_\phi)_{,c,d}\delta^{ac}\delta^{bd}\right)
\ea
$P_\psi,~P_Q$ have been chosen so as to be the true conjugate momenta to $\phi,~Q$.
 Ie, at second order
\beq
\int \pi^{ab}\dot \gamma_{ab}=e^2 \int(P_\phi\dot \phi +P_Q \dot  Q)d^3xdt
\eeq
In addition, I choose
\beq
N=1+e\alpha~~;~~~~~~~~~
N_a= e\beta_{,a}
\eeq
In the following, I will assume that all spatial dependence is of the form
$e^{\pm ik\cdot x}$ so $\nabla^2$ becomes $-k^2$.
Substituting these into the action, and examining  only 
 terms at second order in $e$, one gets  
\beq
S= \int P_\phi\dot \phi+P_Q \dot  q +P_\psi \dot \psi 
-\alpha H_{0(1)} +\beta H_{(1)}^a~_{,a} - H_{0(2)} d^3xdt
\eeq
where the   $()$ index indicates the order (in powers of $e$) of the term 
in question. Although I could write out the detailed form of the various
 terms, it would not be illuminating at this stage.
To mimic the longitudinal gauge choice, I will choose my coordinates so 
that $Q=\dot  Q=\beta=0$. From the variation equation of the action with
 respect to $P_Q$, this  results in the relation $P_Q={1\over 6}k^2 P_\phi$.
 Using the Hamiltonian equations for $\dot  P_Q$ and$\dot  P_\phi$,
 derived by varying the action with respect to $Q$ and $\phi$, one 
finds that this choice of gauge leads to the relation  $\alpha=-\phi$.

The resulting constraint equations are
\ba
0=H_{0(1)}=&&-{1\over 2 \dot  H}\left( -\dot \Psi_0(a^3\ddot H  +6a^3 H\dot  H P_\psi)\psi
\right.\nonumber\\
&&\quad\quad\quad\quad \left. +8a^3(3\dot  H^2+3H^2\dot  H-\dot  H k^2)\phi
+2H\dot  H P_\phi +2\dot \Psi_0 \dot  H \right)\\
0=H_{(1),a}^a=&& -k^2\left({1\over 3a^2}P_{\phi}+4Ha\phi -a \dot \Psi_0\psi\right)
\ea
I then solved these for $P_\psi$ and $\psi$, and substituted the 
results back into the action. Clearly, the terms (the constraints)
multiplying  $\alpha$ and $\beta$ vanish, and the action arises 
solely from the symplectic form and $H_{0(2)}$ terms. The result 
\ba
S= &&\int \left[{4k^2\over 16\dot H a^2} P_\phi \dot\phi +{1\over 72\dot H a^5}k^2\left( \dot H P_\phi^2 
 + (48a^3 H\dot H + 24 a^3 \dot H  ) P_\phi \phi 
\right.\right. \nonumber \\
\quad\quad\left.\left. +(288 a^6 H^2 \dot H +144 a^6 H \ddot H +144k^2 a^4 \dot
H)\phi^2\right)\right]dt
\ea
was still somewhat of a mess, so I   
  defined the new variable $w$ 
 by 
\beq
\phi={w H\over a}.
\eeq
Also, let I had to  change define the momentum,$P_\phi$, (which 
we note is no longer conjugate to $\phi$) to a new one,
 $P_w$, which was  conjugate to $w$ 
(ie, the momentum is conjugate if  the action contains the term $P_w\dot  w$ with unit 
coefficient). The necessary transformation is
\beq
P_w= -{H\over 3 a^3\dot  H }k^2P_\phi
\eeq
The action finally reduces to
\beq
S= \int \left[P_w\dot  w +{\dot  H a\over 8 H^2}P_w^2 +{{\dot  H+H^2}\over H}P_w w +
     2H^2k^2\left({2a H^2\dot  H 
     +a^2H\ddot H +k^2\dot  H \over a^3 \dot  H^2} \right)w^2\right] dt
\eeq
Making another series of transformations
\ba
w&&= W{\sqrt{- \dot  H a}\over 2 kH}\\
P_w&&= \left((P_W-w\left[{\ddot H\over 2\dot  H}
+{3\over 2} H\right]\right) {kH\over \sqrt{-\dot  H a}}
\ea
  and performing the  appropriate temporal integrations by parts 
 (in order to remove terms containing $W\dot P_W$ or $W\dot W$)
we finally get
\beq
S= P_W\dot W -\half\left(P_W^2 +\left({k^2\over a^2}
-{\sqrt{-\dot  H a}\over H}\partial_t^2 \left({H \over\sqrt{- \dot  H a}}\right)\right)W^2\right)
\eeq
Note that this is just the Hamiltonian for a Harmonic oscillator
 with time dependent frequency. I could further reduce the action
 by changing time to conformal time $\tau$, and  defining a new 
variable ${\cal W}=W/a$. This would remove the $1\over a^2$ 
 dependence in the  term proportional to $k^2$, but the system 
is simple enough as it stands.  

This Hamiltonian action can be quantized in the usual manner. Assuming 
that at very early times the state for this variable is in the ground
 state for this Hamiltonian, one obtains that it is be the variable 
$ W$ which will have the vacuum quantum amplitude of fluctuations, 
which, using the WKB approximation  at early times,   
gives $|W|\approx \sqrt{a\over2 k}$ at those times. 

Furthermore one can solve the equation for $w$ exactly if one makes
 certain assumptions about the background solution.
Varying the action with respect to $w$ and $P_w$ and eliminating 
$P_w$ gives an equation for $w$:
 \beq
\ddot w -\left(H+{\ddot H\over \dot  H} -2{\dot H\over H}\right)\dot  w +{k^2\over 
a^2} w=0 \label{w-eqn}
\eeq

For long wavelengths, using our known solutions for $\phi$, the 
solutions for $w$ is $-\epsilon -\beta \int a(t)dt +\beta a/H$. 
(Substituting this into this equation for $w$ verifies it as the 
general solution.) We now need to match this to the solutions for
 large k (or small time),   using the now known amplitude for $W$ 
at early times. The usual method is to match assuming that one is in De
Sitter space-time at early times. However,I find this an uncomfortable
 procedure, as the equation for $w$, and   the relation between
$w$ and $W$, the simple quantum variable,  is   singular as $H$ 
goes to a constant, since both depend on $\dot H$ and $\ddot H$. Furthermore,
 both the equations for $w$ and $W$ are potentially singular in this limit.
 Instead I will assume that 
\beq
  a(t)=t^\mu,
\eeq
with $\mu $ a large
constant. (The limit, $\mu\rightarrow \infty$, corresponds to De Sitter 
space.)  Under this assumption for $a$, the equation for $w $ becomes
\beq
\ddot w  - {\mu\over t} \dot  w +{k^2\over t^{\mu}}w=0
\eeq
Defining the new variable $\tau= -\int {dt\over a}= {1\over (\mu-1)t^{\mu-1}}$, 
the conformal time, the equation for $w$ becomes
\beq
\partial_\tau^2 w  +2{\mu\over(\mu-1) \tau}\partial_\tau w  +k^2 w =0
\eeq
which has as solutions Bessel functions
\beq
w = (k\tau)^{-{\mu+1\over 2(\mu-1)}} \left(A~~ J({\mu+1\over 2(\mu-1)},k\tau) 
+ B~~ Y({\mu+1\over 2(\mu-1)},k\tau)\right)
\eeq
The solution for $W$ is then
\beq
W= {k\sqrt{{\mu}\over a}}w=k \sqrt{\mu} ((\mu-1)\tau)^{\mu\over 2(\mu-1)} w
 \eeq
Now, as argued above,$ W$    will have amplitude of order 
$\sqrt{a/2k}$ at early times. This gives the equation for $A,~B$ of
\beq
A=B={1\over \sqrt{2\mu}} k^{\mu-3\over 2(\mu-1)} (\mu-1)^{-{\mu\over\mu-1}}
\eeq
($A,~B$ are actually quantum operators, and these expressions are
 shorthand for $\sqrt{<A^2>}$ and $\sqrt{<B^2>}$.)
Matching to the solution for long wavelengths (large $t$ or small $\tau$), 
namely
\ba
w&=-\epsilon-\beta  t^{\mu+1}( {1\over (\mu+1)}-{1\over \mu })
=-\epsilon+{\beta\over\mu(\mu+1)}t^{\mu+1} \\
&=A {(k)^{\mu+1\over 2(\mu-1)}\over \Gamma(1+{\mu+1\over 2(\mu-1)})}
 +B {\Gamma({\mu+1\over 2(\mu-1)})\over \pi \tau^{\mu+1\over2(\mu-1)}}
\ea
Thus
\beq
\beta={\Gamma(1/2) \sqrt{\mu}\over k^{3\over 2} \pi}
\left[ 2^{1\over \mu-1}\Gamma({\mu+1\over 2(\mu-1)})(1+{1\over \mu}) 
\over k^{1\over 2(\mu-1)}({\mu-1\over\mu+1})^{2\mu-1\over 2(\mu-1)}
(\mu+1)^{1\over 2(\mu-1)}\Gamma(\half)\right]
\eeq
Since the term proportional to $\epsilon$ dies out rapidly, I will
 not bother with giving its value. .This gives the amplitude for the
 quantum fluctuations during the period while the fluctuations are 
outside the Hubble radius. Note that they depend on $\mu$. We can 
write $\sqrt{\mu}={H\over \sqrt{-\dot  H}}=2{H\over\dot \Psi_0}$. 
The term in square brackets  is a function of $\mu$ which is almost 
unity for large $\mu$, but will give the slow change in the spectrum 
($k$ dependence) as $\dot\Psi_0$ changes slowly during the course of inflation.

In conclusion, {\bf if} one accepts the above procedure for determining 
the initial quantum fluctuations
(an assumption which is not altogether beyond question) and {\bf if} one 
accepts the validity of the 
linear theory for studying the determination of the initial amplitude for 
the fluctuations, then 
the answer which Grishchuk obtains for the final density fluctuations is 
wrong. 
\subsection{Comparison to other Reductions}
As mentioned, such reductions have been carried out previously. It may however be of interest to relate my original approach to this problem. 
Mukhanov, Feldman and Brandenberger \cite {MB} report on such a reduction of the Lagrangian action. 
They derive a reduced Lagrangian action in terms of the longitudinal gauge variable
\beq
v= a (\psi-(\dot\Psi_0/H) \phi)  
\eeq
Using the constraint, $E^{(1)}_{ti}=0$, this reduces to 
\beq
v= -{4H\over a \dot\Psi_0}  \partial_t \left( {a \over H}\phi \right)
\eeq
or, in terms of the reduced variable $w$,
\beq
v= -{2H\over a \sqrt{-\dot H}}\dot w
\eeq
My first reaction to this was to notice that in the long wavelength limit, 
$\dot w$ eliminates the long wavelength solution proportional to 
$\epsilon$ and thus depends only on $\beta$. Thus it would appear that 
the variable $v$ does not capture the full set of physical solutions 
to the perturbation equations in the long wavelength limit,  
making it an unsuitable candidate for quantization. However, the relation is 
more subtle, and in particular it demonstrates another case where 
the $k\rightarrow 0$ limit is not the same as the $k=0$ case. 

Examining  equation \ref{w-eqn}, we notice that it can be 
written as a relation between $w$ and $v$. In particular, we note that
\beq
k^2 w= {aH^2\over \dot H} \partial_t {a\dot H\over H^2}v
\eeq
 Thus, although the zeroth order $\epsilon$ dependent solution for
 $w$ does not contribute to $v$, the next $k^2$ order part of the 
solution does. There is thus a direct relation between the two 
linearly independent  solutions for $w$ and the two  for $v$, 
despite the apparent contradiction when $k=0$.

\section{Higher order Perturbations}
Let us now return to the question of whether or not the higher order long
 wavelength perturbations act to renormalise the cosmological constant, as discussed in ABM.
Since to first order, the solutions locally (when spatial derivatives of 
the
perturbations are neglected) look like simple coordinate transformations 
of the
homogeneous solutions, their effect on the metric to higher order will 
again be
that of simple coordinate transformations. Ie, in the long wavelength limit, 
the 
local effect of the perturbations will just be identical to a simple coordinate 
transformation of the background equations. As far as the local physics 
is concerned, the 
evolution of the universe is identical to that of a homogeneous universe, 
with 
unrenormalised coupling constants. I.e.,the effective cosmological constant 
will not
be renormalized as far as local, sub-Hubble radius physics, is concerned.

However, their analysis is not concerned with local physics. Rather what 
ABM   
 argue  is that, if we examine the average 
evolution of the universe in the large,  these long wavelength perturbations 
act to alter the effective long wavelength evolution of the universe as a  
negative cosmological constant would. To examine this let us, again in the longitudinal 
gauge, examine the behaviour of the long wavelength perturbations. However, 
I will not follow their technique. In their approach, they wish to regard 
the first order perturbations as contributing an extra stress energy tensor 
to the zeroth order equations of motion (ie,   the effective stress 
energy tensor of the first order equations renormalizes the zeroth order 
stress energy tensor.)  I do not wish to follow this procedure as it raises 
delicate problems in consistency.
Since by hypothesis, the zeroth order equations do not obey the zeroth order 
Einstein equations, a consistent derivation of the equations of motion 
becomes difficult.  Instead I will follow the procedure of consistently 
(though probably not convergent) expanding in a series of small perturbations. 
Ie, the metric will be assumed to have the form of
\ba
g_{\mu \nu}= g_{0 \mu\nu} + e g_{1\mu\nu}+e^2 g_{2\mu\nu}+...\\
\Psi=\Psi_0 +e \psi_1+e^2  \psi_2+...
\ea
where the equations of motion are then consistently expanded as a function 
of $e$. The effective stress energy tensor of the first order 
perturbations will act as the source for the second order perturbations in the 
metric and field.  

The biggest problem is that contribution to the second order terms from 
the first
 order perturbations (ie the effective energy momentum tensor for the second
 order perturbations) depend on the gauge chosen for the first order terms.
 Assume that we have two different gauge invariant formulations, determined
 by two separate vectors $X^\mu$ and $\bar X^\mu$. 

  If in one system, the second 
order
 perturbations are given by
$ \delta g_{(2)\mu\nu}$, then in the other gauge, the solutions will be (see
appendix A)
\beq
\delta\bar g_{(2)\mu\nu} = \delta g_{(2)\mu\nu}  + 2\pound_\Delta \delta 
g_{(1)(\mu\nu}
 +  \pound_\Delta\pound_\Delta g_{(0)\mu\nu} + \pound_{\Delta_2} 
g_{(0)\mu\nu}
\eeq
where $\Delta$ is the (first order) vector field which transforms the first 
gauge 
fixing to the second. $\Delta_2$ is the second order part of this fixing.
(note that in most of the work in ABM, was only concerned with the first 
order
 gauge fixing vectors. Since the second order equations were never solved, 
the
 need for the second order components of the $X^{\mu}$ was not there.)

As I show in the appendix, by an appropriate choice of the first
 and second  order gauge
fixing, I can set the second order homogeneous perturbation 
in the metric to whatever value I wish.

Let me be more definite. Let us first work in the ABM gauge, the longitudinal
gauge.

Going back to the observation that the first order solutions are, locally, 
just coordinate
 transformations, the reason for the apparent re-normalization of the cosmological 
constant
 is thus also clear. At each point the universe acts identically to the way it acts at each 
other point, except
 displaced in time by an amount which varies from place to place. If one 
now averages the
 universe over a fixed time slice, the averaged value of the expansion 
at fixed time will
 not be the same as the expansion rate at the averaged value of the time 
because of 
the non-linear nature of the expansion with time. Naively one would expect, 
$<a(\tau(t,x)>=a(<\tau(t,x)>) + a''(<\tau(t,x)>)(<\tau(t,x)^2>-<\tau(t,x)>^2)$,
 where $\tau$ is the uniform time (naively the time defined such that $\delta\psi$ 
is zero along the $\tau$ constant surfaces). The second term is thus present 
because 
of the non-linear relation between $a$ and $t$.

These conclusions about the  also seem to be in agreement with the 
analysis of Salopek \cite{Salopek} who uses the Hamilton Jacobi 
methods to exactly solve for the evolution of the universe in the 
long wavelength limit, for restricted choices of the matter fields. 
His results seem to me also to suggest that the long wavelength 
evolution does not renormalise the cosmological constant. 

After completion of this work, the paper by Kodama and Hamazaki\cite{Kodama}
 was brought to my attention where they have found the solution to the
 long wavelength equations for a general multicomponent system. They
 also find that the long wavelength limit does not correspond to the 
homogeneous solutions for the scalar perturbations.

\section*{Appendix A}

To derive the equation for the perturbations, let us first define what we mean
by the expansion of the metric in the various orders. Consider the space of all
solutions of Einstein's equations. This space can be defined by the set of
tensor components $g_{\mu\nu}(x)$ and fields $\phi(x)$. Now consider a path
through this space of all solutions parameterized by an arbitrary parameter
$e$. Thus we will have a set of functions $g_{\mu\nu}(x,e)$, and
we will choose the path and $e$ such that at $e=0$, the metric is
the given background metric. For each value of $e$ these are solutions of
the equations, and thus we have
\beq
G_{\mu\nu}\left( g_{\alpha\beta}(x,e)\right)= 
T_{\mu\nu}( g_{\alpha\beta}(x,e),\phi(x,e)).
\eeq 
Since this equation is valid by assumption for all $e$, derivatives of
these equations with respect to $e$ will also be satisfied. Furthermore,
assuming that these equations are analytic in $e$ near $e=0$, we
can also create a power series expansion of the set of solutions along the path
in $e$. 

Now consider another path through this space of solutions defined by an
$e$ dependent coordinate transformation 
\beq
 \tilde x^\alpha= \tilde x^\alpha(x,e).
 \eeq
 such that 
 \beq
 g_{\mu\nu}(x,e) = {\partial \tilde x^\alpha \over \partial x^\mu}
 {\partial \tilde x^\beta \over \partial x^\nu}
 \tilde g_{\mu\nu}(\tilde x(x,e),e) \label{gg}
 \eeq
 for some other path through the space of solutions defined by 
 $tilde g_{\mu\nu}(\tilde x,e)$.
 
 Now define 
 \beq
 \eta^\mu= {\partial \tilde x^\alpha(x,e) \over \partial
 e} {\partial X^\mu(\tilde x(x,e),e)\over \partial \tilde
 x^\alpha}
 \eeq
 where $X^\mu(\tilde x,e)$ is defined so that 
 $X^\mu(\tilde x(x,e),e)=x^\mu$ for all $e$. Also define 
 the second order infinitesimal coordinate transformation by   
 \beq
 \zeta^\mu= {\partial\eta^\mu(x,e)\over \partial e}
 \eeq
 Taking the second derivative of the equation \ref{gg} with respect to
 $e$ at $e=0$, and assuming that $\tilde x^\alpha(x,0)=x^\alpha$,
  we obtain the second order coordinate transformations. After some
 algebra, one finds that 
 \beq
 {\partial^2 g_{\mu\nu}(x,e=0)\over e^2}= {\partial \tilde
 g_{\mu\nu}(x,e=0)\over \partial e^2 } +\pound_\zeta  \tilde g_{\mu\nu}(x,e=0) 
 +2\pound_\eta  {\partial\tilde g_{\mu\nu}(x, e=0)\over\partial e}+
 \pound_\eta \pound_\eta \tilde g_{\mu\nu}(x,e=0)
 \eeq
 Ie, the second order metric components are not independent of the first order
 coordinate transformation.
 
 Let us consider the particular case where the first order solution 
 $\tilde h_{\mu\nu} \equiv {\partial \tilde g_{\mu\nu}(x, e=0)\over
 \partial e}$ is evaluated in the longitudinal gauge (ie, is diagonal),
 and has spatial dependence with spatial wave-vector $k$. Assuming that $k$ is
 very small (super-horizon modes), I will neglect terms proportional to 
 $k$ or $k^2$. ( In the ABM paper, the first order $k$ dependent terms would be
 zero because of the assumption that the fluctuations are statistically
 homogeneous, and have no preferred direction). Furthermore, let me assume that
 only $\eta^0$ is non-zero, with arbitrary time dependence and with spatial
 dependence with wave vector $k$. Furthermore, let us look only at the effect of
 this first order gauge transformation on the second order spatially homogeneous
  terms. Then, 
  \beq
  \pound_\eta \tilde g_\mu\nu = 2\dot\eta^0\delta_{0\mu}\delta_{0\nu} + 2a\dot a \eta^0
  \delta_{ij}
  \eeq
  and 
  \beq
  \pound_\eta( 2 \tilde h_{00} + \pound _\eta g_{00} )
  = \eta^0(2\dot {\tilde h_{00}} + 2ddot\eta^0) 
  + 4( \tilde h_{00} +  \dot\eta^0) \dot\eta^0  
  \eeq
  and
  \beq
  \pound_\eta( 2 \tilde h_{ij} + \pound _\eta g_{ij} )
  =2\eta^0 \left( \dot{\tilde h}_{ij}
  +  \dot\eta^0(\eta^0(a\ddot a+\dot a^2) +\dot\eta^0 (a\dot a))\delta_{ij}
  \right)
  \eeq
  Using the gauge choice from the main paper, such that $h_{ij}=2a^2 \Phi
  \delta_{ij}$, $h_{00}= 2 phi$, $\phi=Phi\approx   const$, we have, neglecting
  terms of order $k^2$,
  \ba
  \pound_\eta( 2 \tilde h_{00} + \pound _\eta g_{00} )=&&
  2\eta^0 \ddot \eta^0 +4 \phi \dot\eta^0 +4(\dot\eta^0)^2\\
  \pound_\eta( 2 \tilde h_{ij} + \pound _\eta g_{ij} )=&&
  a^2\delta_{ij} 2\eta^0( 2H\phi + (\dot H +2H^2)\dot \eta^0) + H(\dot\eta^0)^2 
  \ea
  Note that all of the off diagonal terms due to this first order gauge
  transformation are zero because of the averaging over $k$. 
  We know the gauge changes for the homogeneous modes, with 
  $\zeta^0(t)$ and $\zeta^i=\lambda x^i$, which gives the changes
  \beq
  \pound_\zeta \tilde g_{00}= 2\dot\zeta^0\\
  \pound_\zeta \tilde g_{ij}= (-2a\dot a\zeta^0+\lambda)\delta_{ij}
  \eeq
  The second order terms from the first order gauge transformations cannot be
  cast into this form.  One thus has two independent functions of $t$, namely
  that produced by the first order inhomogeneous gauge transformation, and that
  caused by the second order homogeneous gauge transformation. By making appropriate choices of both the first order
  gauge with spatial wavenumber $k$ and of the homogeneous second order gauge,
  and neglecting terms of order $k^2$,
  one can set the two second order homogeneous terms in the metric (ie the $00$
  and the diagonal spatial components) to any value one
  desires, including 0.

\section*{Acknowledgements}

I would like to thank R. Brandenberger for discussion and for
 bringing the issue to my attention. I would also like to thank S. Mukhanov 
for extensive email in which he corrected my misconception about the relation
 between the MFB reduction and the one given in this paper.  L. 
Grishchuk and I also have had extensive email discussions on the issues
 raised here. 
Finally I  thank the Canadian Institute for Advanced Research
 and the Natural Sciences and Engineering Research Council for 
support during this work. The algebraic manipulations necessary
 in driving the equations and in reducing the Hamiltonian action 
were done with the help of GRTensorII under MapleV4.

\end{document}